\input amstex
\magnification 1200
\TagsOnRight
\NoBlackBoxes
\baselineskip 20 pt
\parskip 8 pt

\centerline {\bf EXPLICIT SOLUTIONS TO}
\centerline {\bf THE KORTEWEG-DE VRIES EQUATION}
\centerline {\bf ON THE HALF LINE}

\vskip 10 pt
\centerline {Tuncay Aktosun}
\vskip -8 pt
\centerline {Department of Mathematics}
\vskip -8 pt
\centerline {University of Texas at Arlington}
\vskip -8 pt
\centerline {Arlington, TX 76019-0408, USA}

\centerline{Cornelis van der Mee}
\vskip -8 pt
\centerline{Dipartimento di Matematica e Informatica}
\vskip -8 pt
\centerline{Universit\`a di Cagliari}
\vskip -8 pt
\centerline{Viale Merello 92, 09123 Cagliari, Italy}

\noindent {\bf Abstract}:
Certain explicit solutions to the Korteweg-de
Vries equation in the first quadrant of the $xt$-plane
are presented. Such solutions involve
algebraic combinations of truly elementary functions, and their
initial values correspond to rational reflection
coefficients in the associated Schr\"odinger equation.
In the reflectionless case such solutions reduce to
pure $N$-soliton solutions.
An illustrative example is provided.


\vskip -8 pt
\par\noindent {\bf Short title:} Explicit solutions to the KdV equation
\newpage

\noindent {\bf 1. INTRODUCTION}

Consider the celebrated Korteweg-de Vries (KdV) equation
$$\displaystyle \frac{\partial u}{\partial t}+\eta \,\frac{\partial u}{\partial x}
-6u\,\frac{\partial u}{\partial x}
+\frac{\partial^3 u}{\partial x^3}=0,
\tag 1.1$$
where $x$ and $t$ denote the spatial and temporal variables, respectively,
and $\eta$ is a nonnegative constant [1,2] that can be chosen as $0$ or $1.$
The KdV equation is used to model [3-5] the propagation of water waves in
long, narrow, shallow channels; it also arises in other areas such as
hydromagnetic waves in a cold plasma, ion-acoustic waves, and
acoustic waves in harmonic crystals.

The KdV equation is one of the most well-known and most widely analyzed
nonlinear partial differential equations. It has
many remarkable aspects [4,5]. For example,
it possesses traveling wave solutions known as solitons.
The numerical studies on the KdV equation by Zabusky and Kruskal [3]
led to the discovery of multi-soliton solutions,
where the individual solitons interact nonlinearly at close
distance and then move away from each other without changing their
shapes. In their celebrated paper [6] Gardner, Greene, Kruskal, and Miura showed
that the initial-value problem for the KdV equation can be
solved via the ``inverse scattering transform"
associated with the Schr\"odinger equation.
This led to the discovery that certain nonlinear partial differential equations
are ``completely integrable": They can be solved via an inverse
scattering transform and they have some interesting common properties
such as possessing soliton solutions, Lax pairs, and infinitely many
conserved quantities.

A pure $N$-soliton solution to the KdV equation can be written explicitly as [4,5,7,8]
$$u(x,t)=-2\,\displaystyle\frac{\partial^2 \log\left( \det \Gamma(x;t)\right)}
{\partial x^2},\tag 1.2$$
where $\Gamma(x;t)$ is the $N\times N$ matrix whose $(j,l)$ entry is given by [9]
$$\Gamma_{jl}=\delta_{jl}+\displaystyle\frac{c_j\,
e^{-2\kappa_j x+8\kappa_j^3 t+2\eta \kappa_j t}}{
\kappa_j +\kappa_l},
\qquad 1\le j,l\le N,\tag 1.3$$
with $\delta_{jl}$ denoting the Kronecker delta, the
$\kappa_j$ are $N$ distinct positive constants corresponding to the
bound states of $u(x,0),$ and
the $c_j$ are $N$ positive constants known as the bound-state
norming constants. Pure soliton solutions
are trivial in the sense that the potential $u(x,0)$
corresponds [3-5] to a zero reflection
coefficient in the Schr\"odinger equation.
There are many important ways explicit solutions to the KdV equation may help
us to understand nonlinearity better.
For example, it is of great importance [1,10] to determine function spaces
containing the initial data $u(x,0)$ so that $u(x,t)$ is globally
well behaved (i.e. does not blow up during $t\in[0,+\infty)$) or
only locally well behaved (i.e. remains finite during
$t\in[0,\tau)$ and blows up at some finite time $\tau$), and
explicit solutions may help us to understand the global or local
well-posedness of initial-value problems for the KdV equation.
Explicit solutions may
contribute to the development or improvement of numerical methods for (1.1), and
they may also be useful
to check the accuracy of existing numerical methods.

In this paper we present a method leading to certain explicit solutions to
the KdV equation in the first quadrant of the $xt$-plane.
Let us emphasize that our aim is not to solve the initial-boundary-value problem for (1.1)
in the quarter plane. Instead, we are interested in producing some explicit solutions to
(1.1) in terms of truly elementary functions. We produce certain
explicit solutions to (1.1) having the form
(1.2), where $\Gamma(x;t)$ is the matrix appearing in (4.1). From
(4.1) it is seen that $\Gamma(x;t)$ can be constructed explicitly by specifying
a constant $P\times P$ matrix $A,$ a constant $P$-row vector $C,$ and
a constant $P$-column vector $B.$ We also show that such solutions can equivalently
be written as in (4.9). In fact, it is straightforward (but tedious) to verify that
the right-hand side in (4.9) is a (formal) solution to (1.1) no matter how
$A,$ $B,$ and $C$ are chosen. However, as seen from (4.1),
an arbitrary choice for $A$ may not guarantee
the convergence of the integral in (4.1); even if $\Gamma(x;t)$ obtained from (4.1) exists by
choosing $A$ appropriately, that
particular choice for $A$ and arbitrary choices for $B$ and $C$
may not assure the existence of $\Gamma(x;t)^{-1}$ appearing in (4.9) for
$x\in[0,+\infty)$ and $t\in[0,\tau)$ for some $\tau>0$ or $\tau=+\infty.$
One of our tasks in this paper is to indicate how we may choose $A,$ $B,$ $C$ in order
to assure the existence of $\Gamma(x;t)$ and the positivity
of its determinant for all $x\in[0,+\infty)$ and $t\in[0,\tau),$
which in turn assures the existence and well-posedness of the solution $u(x,t)$ given in (4.9).

One set of possible choices for $A,$ $B,$ $C$
corresponds to the initial values $u(x,0)$ for
$x\in[0,+\infty)$ in such a way that $u(x,0)$ becomes the potential
belonging to a certain class in the one-dimensional
Schr\"odinger equation. For example, $u(x,0)$ may
be viewed as a fragment of a real-valued and integrable
potential, which
has a finite first moment and which
corresponds to a rational reflection coefficient. Other choices may be possible;
e.g., we may further require that the corresponding one-dimensional
potential $u(x,0)$ vanish identically for $x<0.$ In fact, in Section 3 we
outline how $A,$ $B,$ $C$ can be explicitly constructed from such a potential.
All such choices guarantee the existence of the integral in (4.1)
for all $x\in[0,+\infty)$ and each fixed $t.$ This is because such choices,
as a result of using (3.2), assure that each eigenvalue of $A$ has a positive
real part. Hence, the matrix $\Gamma(x;t)$ defined in (4.1) exists
for all $x\in[0,+\infty)$ and each fixed $t,$ its determinant
$\det \Gamma(x;t)$ is continuous in $t$ for every $x\in[0,+\infty),$ and
also $\det \Gamma(x;t)\to 1$ as $x\to+\infty$ for each fixed $t.$
Furthermore, for such choices it
is already known [11] that the resulting $u(x,0)$ is analytic in $x$ and
$\det \Gamma(x;0)>0$ for all $x\in[0,+\infty).$
Consequently, we have $\det \Gamma(x;t)>0$ for all $x\in[0,+\infty)$ and $t\in[0,\tau)$
for some positive $\tau.$ It is remarkable that, for certain choices of $A,$ $B,$ $C,$
we can have $\tau=+\infty,$ as we see from the example in Section 5.

As seen from the analysis in Section 3, in case the
relevant reflection coefficient is zero, our solution $u(x,t)$ reduces
to the pure $N$-soliton solution given in (1.2)-(1.3). This is equivalent to
choosing in (4.1) and (4.9) the matrix $A$ as the $N\times N$ diagonal matrix with $\kappa_j$
appearing in the $(j,j)$ entry, $B$ as the column $N$-vector
having the number $1$ in each entry, and $C$ as the row
$N$-vector with $c_j$ appearing in the $j$th entry.

This paper is organized as follows. In section 2 we mention some of the other methods to solve the
KdV equation and give a brief comparison. In Section 3 we provide a physical motivation
for the derivation of our solutions and show how they may be related
to some scattering data. In Section 4 we show how our solutions can be obtained
by solving the Marchenko integral equation. Finally, in Section 5 we present an
example to illustrate our method.

\noindent {\bf 2. SOME OTHER METHODS FOR THE KdV EQUATION}

As seen from Section 3, our method is based on using the inverse scattering transform, exploiting
the degeneracy of the kernel of the Marchenko integral equation as indicated in
(4.7), and solving the Marchenko equation (4.4) algebraically. There are also
methods to solve the KdV equation without using the inverse scattering transform.
For example, the technique [12,13] based on using the B\"acklund transformation,
the technique [8,14] using the Darboux-Crum transformation, the Wronskian techniques
and their generalizations [7,15-18], and the Hirota method [19].
Such methods are also used to produce certain exact solutions to the
KdV equation. The idea behind the methods using the transformations of B\"acklund
and Darboux-Crum is to obtain new solutions to the KdV equation from other previously known
solutions. The basic idea behind the Wronskian methods and the Hirota method is to represent
the solution to the KdV equation in a particular form so that certain functions in the
representation satisfy certain linear differential equations even though the solution itself
satisfies a nonlinear differential equation. The explicit solutions produced by our method
have the same representation (1.2) or (4.2) as in the Wronskian methods; however, our matrix
$\Gamma(x;t)$ (or a part of it) does not necessarily satisfy a linear partial differential
equation as expected in the
Wronskian methods. In the method based on the Darboux-Crum transformation, the solution to
the KdV equation has the same representation as in (1.2)
or (4.2), provided the initial solution is chosen as zero;
there is certainly some connection between that method
and our method because they both yield the $N$-soliton
solution in the easiest case; however, any possible connection in the more general case is not
clear at the moment and requires a detailed analysis, which we plan to do in the future.
For the time being, we only emphasize that our exact solutions satisfy the half-line
KdV equation with the drift term $\eta u_x$ where we can choose $\eta=0$ or
$\eta>0$ at will, they include some global-in-time solutions as well as some local-in-time solutions,
and they are algebraic combinations of truly elementary functions.
One advantage of our method
is that it can be generalized to obtain certain explicit solutions to the matrix KdV equation
as well as to the scalar and matrix nonlinear
Schr\"odinger equations.

Some other explicit solutions to the KdV equation known in the literature include
algebraic solitons [20-22], rational solutions [22,23], various singular solutions [24-26]
such as positons and negatons,
solutions [22] to the periodic and other KdV equations,
solutions [27] that are not quite as explicit but expressed in terms of certain
projection operators, and various other solutions [28,29]. It is already known that
some rational solutions can be obtained by letting the bound-state energies go to zero
in the $N$-soliton solutions. We plan to do in the future a detailed
comparison between our solutions (and
their possible generalizations) and exact solutions obtained by other methods. Some
generalizations of our solutions might be obtained by letting the dimension of
the matrix $A$ in (3.3) go to infinity, by choosing the entries
of $A$ and $C$ given in (3.3) in some particular way or by letting some entries go
to certain limits such as zero, and by analyzing the singularities encountered at $t=\tau.$

\noindent {{\bf 3. SOME POSSIBLE CHOICES FOR} $A,$ $B,$ $C$}

In this section we indicate a possible set of choices for $A,$ $B,$ $C$
appearing in $\Gamma(x;t)$ of (4.1) so that
the resulting function $u(x,t)$ given in (1.2) or (4.2), or equivalently that in (4.9),
is an explicit solution to (1.1) for all $x\in[0,+\infty)$ and $t\in[0,\tau)$
with some positive $\tau.$

Starting with the initial value $u(x,0)$ with
$x>0,$ we extend it to the whole line by choosing $u(x,0)\equiv 0$ for
$x<0$ and we uniquely determine the corresponding
scattering data $\{R,\{\kappa_j\},\{c_j\}\}.$ Here,
$R(k)$ is the corresponding right reflection coefficient [30-34],
the set of constants $\kappa_j$ with $0<\kappa_1<\dots<\kappa_N$ corresponds to the
bound states associated with the full-line potential $u(x,0),$ and
the set of constants $c_j$ corresponds to the associated bound-state norming constants.
The construction of $\{R,\{\kappa_j\},\{c_j\}\}$ can be accomplished through
the following steps:

\item{(a)} Given $u(x,0)$ for $x\in[0,+\infty),$
uniquely determine the
corresponding Jost solution $f_{\text r}(k,x)$ from the right by solving
the initial-value problem for the half-line Schr\"odinger equation
$$\displaystyle\frac{d^2 f_{\text r}}{dx^2}+k^2 f_{\text r}=u(x,0)\,f_{\text r};
\qquad f_{\text r}(k,0)=1,\quad
\displaystyle\frac{d f_{\text r}(k,0)}{dx}=-ik.$$

\item{(b)} Recover the corresponding right reflection coefficient
$R$ and the transmission coefficient $T$ with the
help of the asymptotics [30-34] of
$f_{\text r}$ as $x\to+\infty,$ namely by using
$$f_{\text r}(k,x)=\displaystyle\frac{1}{T(k)}\,e^{-ikx}
+\displaystyle\frac{R(k)}{T(k)}\,e^{ikx}+o(1),
\qquad x\to+\infty.$$
It is known [30-34] that $T$ is
related to $R$ via
$$T(k)=\displaystyle\prod_{j=1}^N
\left(\displaystyle\frac{k+i\kappa_j}{k-i\kappa_j}
\right)\,\exp\left(\displaystyle\frac{1}{2\pi i}
\int_{-\infty}^\infty ds\,
\displaystyle\frac{\log(1-|R(s)|^2)}{s-k-i0^+}\right),\qquad k\in\overline{\bold C^+},\tag 3.1$$
where $\overline{\bold C^+}:=\bold C^+\cup\bold R,$
$\bold C^+$ is the upper half complex plane, and the $0^+$ indicates that the limit from
$\bold C^+$ should be used to evaluate $T(k)$ for real $k$ values.

\item{(c)} Construct the set $\{\kappa_j\}_{j=1}^N$ by using (3.1).

\item{(d)} Construct the set of positive constants $\{c_j\}_{j=1}^N$ by using [30]
$$c_j=-[\text{Res}(T,i\kappa_j)]^2
\left[\displaystyle\frac{1}{2\kappa_j}+\displaystyle\int_0^\infty dx\, f_{\text r}(i\kappa_j,x)^2
\right],$$
where the purely imaginary constant $\text{Res}(T,i\kappa_j)$ denotes the residue
of $T$ at $k=i\kappa_j.$

Having constructed $R$ which is a rational function of $k,$
we determine all its poles in $\bold C^+$ and the coefficients in the
partial fraction expansion
of $R$ at such poles. It is known [30-34] that
$R(-k^*)=R(k)^*$ with the asterisk denoting
complex conjugation, and hence such poles are either located on the
positive imaginary axis $\bold I^+$ or they occur in pairs
symmetrically located with respect to $\bold I^+.$ Let us use
$M$ to denote the number of poles in $\bold C^+$ without counting the multiplicities,
and let us order them
in such a way that the first $n$ pairs are located off $\bold I^+$
at $k=\pm\alpha_j+i\beta_j$ with $\alpha_j>0$ and
$0<\beta_1\le \dots\le \beta_n;$ in case several distinct $\alpha_j$ values correspond to the
same $\beta_j,$ we can further arrange $\alpha_j$ in increasing order. We
choose our notation so that the remaining $M-2n$ poles occur at $k=i\omega_j$ on
$\bold I^+$ with
$0<\omega_{2n+1}<\dots<\omega_M.$ We let $m_j$ indicate the
multiplicity of the $j$th pole.

Let $\Pi R$ denote the part of the partial fraction
expansion of $R$ containing only the poles in $\bold C^+.$
We have
$$\Pi R(k)=\displaystyle\sum_{j=1}^{n}\displaystyle\sum_{s=1}^{m_j}
\left[\displaystyle\frac{(-i)^s(\epsilon_{js}+i\gamma_{js})}
{(k-i\beta_j -\alpha_j)^s}+\displaystyle\frac{(-i)^s(\epsilon_{js}-i\gamma_{js})}
{(k-i\beta_j +\alpha_j)^s}\right]+
\displaystyle\sum_{j=2n+1}^{M}\displaystyle\sum_{s=1}^{m_j}
\displaystyle\frac{(-i)^s r_{js}}
{(k-i\omega_j)^s}.\tag 3.2$$
As a result of $R(-k^*)=R(k)^*,$ the constants $\epsilon_{js},$
$\gamma_{js},$ and $r_{js}$ appearing in (3.2) are all real; in fact, we have
$$\epsilon_{js}+i\gamma_{js}=\displaystyle\frac{i^s}{(m_j-s)!}
\displaystyle\frac{d^{m_j-s}}{dk^{m_j-s}}\left[
R(k)\,(k-\alpha_j-i\beta_j)^{m_j}\right]\bigg|_{k=\alpha_j+i\beta_j},
\qquad j=1,\dots,n,$$
$$r_{js}=\displaystyle\frac{i^s}{(m_j-s)!}
\displaystyle\frac{d^{m_j-s}}{dk^{m_j-s}}\left[
R(k)\,(k-i\omega_j)^{m_j}\right]\bigg|_{k=i\omega_j},
\qquad j=2n+1,\dots,M.$$

For $j=1,\dots,n,$ let us define $C_j:=2\bmatrix \gamma_{j m_j}& \epsilon_{j m_j}&\dots
&
\gamma_{j 1}& \epsilon_{j 1}\endbmatrix$ and
$$A_j:=\bmatrix \Lambda_j&
-I_2&0&\dots&0&0\\
0& \Lambda_j& -I_2&\dots&0&0\\
0&0&\Lambda_j&\dots&0&0\\
\vdots&\vdots &\vdots &\ddots&\vdots&\vdots\\
0&0&0&\dots&\Lambda_j&-I_2\\
0&0&0&\dots&0&\Lambda_j\endbmatrix
,\quad
B_j:=\bmatrix 0\\
\vdots\\
0\\
1\endbmatrix,$$
where $I_2$ denotes the
$2\times 2$ unit matrix, each column vector
$B_j$ has $2m_j$ components, each $A_j$
has size $2m_j\times 2m_j,$
and each $2\times 2$
matrix $\Lambda_j$ is defined as
$$\Lambda_j:=\bmatrix \beta_j&\alpha_j
\\ -\alpha_j& \beta_j\endbmatrix.$$
Similarly, for $j=2n+1,\dots,M,$ let
$$A_j:=\bmatrix \omega_j&
-1&0&\dots&0&0\\
0& \omega_j& -1&\dots&0&0\\
0&0&\omega_j& \dots&0&0\\
\vdots&\vdots &\vdots &\ddots&\vdots&\vdots\\
0&0&0&\dots&\omega_j&-1\\
0&0&0&\dots&0&\omega_j\endbmatrix,
\quad
B_j:=\bmatrix 0\\
\vdots\\
0\\
1\endbmatrix,
\quad C_j:=\bmatrix r_{j m_j}& \dots
&
r_{j 1}\endbmatrix,$$
where
each column vector
$B_j$ has $m_j$ components and each $A_j$
has size $m_j\times m_j.$
Note that we can write (3.2) as
$$
\Pi R(k)=-i\bmatrix C_1&\dots&C_{M}\endbmatrix
\bmatrix (k-iA_1)^{-1}&0&\dots&0\\
0&(k-iA_2)^{-1}&\dots&0\\
\vdots&\vdots &\ddots&\vdots\\
0&0&\dots&(k-iA_{M})^{-1}
\endbmatrix
\bmatrix B_1\\
\vdots
\\
B_{M}\endbmatrix.$$
The above expression corresponds to a minimal realization [35]
of $\Pi R.$
Associated with the bound-state data $\{\kappa_j,c_j\}_{j=1}^N,$ we let
$$A_{M+j}:=\kappa_j,\quad
C_{M+j}:=c_j,\quad
B_{M+j}:=1,\qquad j=1,\dots,N.$$
Let us also define
$$A:=\bmatrix A_1&0&\dots&0\\
0&A_2&\dots&0\\
\vdots&\vdots&\ddots&\vdots\\
0&0&\dots&A_{M+N}\endbmatrix,\quad
B:=\bmatrix
B_1\\
\vdots\\
B_{M+N}\endbmatrix,\quad
C:=\bmatrix C_1&\dots&C_{M+N}\endbmatrix.\tag 3.3$$
Note that $A$ is a $P\times P$ block square matrix,
$B$ is a column $P$-vector, and $C$ is a row $P$-vector,
where $P$ is the constant given by
$$P:=N+2\sum_{j=1}^n m_j+\sum_{j=2n+1}^M m_j.$$
We also note that all the entries in $A,$ $B,$ and $C$ are real constants.

\noindent {\bf 4. EXPLICIT SOLUTIONS}

In this section we construct our explicit
solutions in terms of the three matrices $A,$ $B,$ and $C.$
In Section 3 we have described
how $A,$ $B,$ $C$ may be related to some scattering data.
Let us define
$$\Gamma(x;t):=I_P+\displaystyle\int_x^\infty dz\, e^{-zA}BC e^{-zA}e^{8tA^3+2\eta At},\tag 4.1$$
where $I_P$ is the $P\times P$ unit matrix. Our main result is that the quantity $u(x,t)$
given as
$$u(x,t)=
-2\displaystyle\frac{\partial}{\partial x}\left[
\displaystyle\frac{\displaystyle\frac{\partial}{\partial x} \det \Gamma(x;t)}
{\det \Gamma(x;t)}
\right],\tag 4.2$$
is a solution to (1.1) as long as $\det \Gamma(x;t)>0$
or, equivalently, as long as the matrix $\Gamma(x;t)$ is invertible.
It is known [11] that $\det \Gamma(x;0)>0$ for $x\in[0,+\infty).$ As seen from (4.1), the
matrix $\Gamma(x;t)$ can be explicitly constructed from
$A,$ $B,$ and $C,$ and as argued in Section 1
we have $\det \Gamma(x;t)>0$
for all $x\in[0,+\infty)$ and $t\in[0,\tau)$ for some $\tau>0.$
There are two possibilities: If $\tau=+\infty$ then the solution $u(x,t)$ given
in (4.2) is a global-in-time solution to (1.1); otherwise, it is a local-in-time solution.

The proof that (4.2) satisfies (1.1) when $\Gamma(x;t)$ is invertible can
be outlined as follows. The solution to (1.1) via the inverse scattering transform
is obtained as in the diagram
$$\CD \{R(k),\{\kappa_j\},\{c_j\}\}
@<<\text{direct scattering}< u(x,0)\\
@V{\text{time evolution}}VV @VV{\text{solution to KdV}}V \\
\{R(k)\, e^{8ik^3t-2i\eta kt},\{\kappa_j\},\{c_j
\,e^{8\kappa_j^3 t+2\eta \kappa_j t}\}\} @>\text{inverse scattering}>> u(x,t)
\endCD\qquad \qquad \quad \tag 4.3$$
The inverse scattering step in (4.3) for $x>0$ can be
accomplished by solving the time-evolved Marchenko equation [4,5,9]
$$\displaystyle K(x,y;t)+\Omega(x+y;t)+\displaystyle\int_x^\infty
dz\,K(x,z;t)\,\Omega(y+z;t)=0,\qquad y>x>0,\tag 4.4$$
where the Marchenko kernel $\Omega(y;t)$ is given by
$$\Omega(y;t):=\displaystyle\frac{1}{2\pi}
\displaystyle\int_{-\infty}^\infty dk\, R(k)\,e^{8ik^3t-2i\eta kt+iky}
+\displaystyle\sum_{j=1}^Nc_j\,e^{8\kappa_j^3t+2\eta \kappa_j t-\kappa_j y}.\tag 4.5$$
If $t=0$ in (4.5) then we
can explicitly evaluate $\Omega(y;0)$ in terms of $A,$ $B,$ $C$ given
in (3.3), and this can be accomplished with the help
of the generalized Cauchy integral formula by using
a contour integration along the boundary of $\bold C^+.$
In general, we cannot evaluate $\Omega(y;t)$ the same way for
all $t>0,$ although there are cases when we might be able do this;
for example, if all the eigenvalues of $8A^3+2\eta A$ have nonpositive real parts,
then we might explicitly evaluate $\Omega(y;t)$ and obtain
$$\Omega(y;t)=C e^{8tA^3+2\eta At-yA}B.\tag 4.6$$
It turns out that the evaluation of (4.5) as (4.6) yields
(4.9), which is a solution to (1.1) as long as
$\Gamma(x;t)$ is invertible. As discussed in Section 1, this invertibility
holds for all $x\in[0,+\infty)$ and $t\in[0,\tau)$ for
some $\tau>0,$ where the value of $\tau$ depends on the value of
$\eta$ and the entries of the constant matrices $A$ and $C$ given in (3.3).
We can write $\Omega(x+y;t)$ as a dot product of a $P$-vector
not containing $x$ and a $P$-vector not containing $y.$ This
separability is easily seen from (4.6) by writing
$$\Omega(x+y;t)=C e^{8tA^3+2\eta At-xA}e^{-yA}B,\tag 4.7$$
where $C e^{8tA^3+2\eta At-xA}$ is a row $P$-vector and
$e^{-yA}B$ is a column $P$-vector. The degeneracy of the
kernel $\Omega(y;t)$ allows us to solve (4.4)
explicitly by algebraic means. In fact, its explicit solution is given by
$$K(x,y;t)=-Ce^{8tA^3+2\eta At-Ax}\Gamma(x;t)^{-1}e^{-yA}B,\tag 4.8$$
where $\Gamma(x;t)$ is the matrix in (4.1). Finally, the time-evolved
potential $u(x,t),$ which is also a solution to (1.1), is obtained
 from (4.8) via [4,5,9]
$$u(x,t)=-2\displaystyle\frac{\partial K(x,x;t)}{\partial x},$$
leading to
$$u(x,t)=2\displaystyle\frac{\partial}{\partial x}\left[Ce^{8tA^3+2\eta At-Ax}
\Gamma(x;t)^{-1}e^{-xA}B\right].\tag 4.9$$
 From (4.1) and (4.9) we obtain
$$u(x,t)=-2\displaystyle\frac{\partial}{\partial x}\,\text{tr} \left[\Gamma(x;t)^{-1}
\displaystyle\frac{\partial}{\partial x}\Gamma(x;t)\right],\tag 4.10$$
where we have used the fact
that in evaluating the trace of a product of two matrices, we can change
the order in the product. Using Theorem~7.3 on p. 38 of
[36], we can write (4.10) also as (1.2) or (4.2).

As indicated in Section 1, it is somehow surprising that any set of
arbitrary choices for $A,$ $B,$ $C$ in (4.1) and (4.9) yields a formal solution
to (1.1).
It can independently and directly be verified that
$u(x,t)$ given in (4.9) is a solution to (1.1) in a region in the $xt$-plane as long as
$\Gamma(x;t)$ exists and is invertible in that region. The verification of this
can be achieved in a straightforward way by taking the appropriate derivatives
of the right-hand side of (4.9) and substituting them in the left-hand side in (1.1).

\noindent {\bf 5. AN EXAMPLE}

We will now illustrate our method by an explicit example.
Consider the scattering data with no bound states and
$$\Pi R(k)=\displaystyle\frac{-2i\epsilon(k-i/2)-\sqrt{3}\gamma}
{(k-i/2)^2-3/4},\tag 5.1$$
where $\epsilon$ and $\gamma$ are some positive constants. Using (5.1)
in (3.3) we obtain
$$A=\bmatrix 1/2&-\sqrt{3}/2\\
\sqrt{3}/2&1/2\endbmatrix,\quad B=\bmatrix 0\\
1\endbmatrix,\quad C=2\bmatrix \gamma&\epsilon\endbmatrix.\tag 5.2$$
Alternatively, we can start with $A,$ $B,$ $C$ given in (5.2)
without even knowing
that they may be related to some scattering data.
The use of (5.2) in (4.1) results in
$$
\aligned
\det\Gamma(x;t)=&1-\displaystyle\frac34\,(\epsilon^2+\gamma^2)
\,e^{2(\eta-8)t-2x}\\
&+\displaystyle\frac12\,
e^{(\eta-8)t-x}\left[
(\sqrt{3}\epsilon-\gamma)\,\sin(\sqrt{3}\eta t-\sqrt{3} x)
+(\epsilon+\sqrt{3}\gamma)\,\cos(\sqrt{3}\eta t-\sqrt{3} x)\right].
\endaligned
\tag 5.3$$
Note that $\det\Gamma(x;t)>0$ for all $x,t\ge 0$ if $(\epsilon^2+\gamma^2)<4/9$
and $0\le \eta\le 8.$
It can directly be verified that
$u(x,t)$ obtained as in (4.2) with $\det\Gamma(x;t)$
given in (5.3) solves (1.1) and hence it is a global-in-time solution. Not imposing
such restrictions on $\epsilon,$ $\gamma,$ and $\eta,$  we still obtain solutions to (1.1),
which may however be only locally well behaved or may even have singularities.

For example, by choosing $\epsilon=\gamma=1/2$ and $\eta=1,$ we obtain the
explicit solution to (1.1) in the form
$$u(x,t)=\displaystyle\frac{\phi(x,t)}{\left[1-\displaystyle\frac38 e^{-2(x+7t)}
+\displaystyle\frac{1}{\sqrt{2}} e^{-(x+7t)}\cos(\sqrt{3}(x-t)+\pi/12)\right]^2},$$
where we have defined
$$\phi(x,t):=6e^{-2(x+7t)}-4\sqrt{2}e^{-(x+7t)}\sin(\sqrt{3}(x-t)-\pi/12)
-\displaystyle\frac{3}{\sqrt{2}}
e^{-3(x+7t)}\sin(\sqrt{3}(x-t)+\pi/4).$$
This solution is valid for all $x\in[0,+\infty)$ and $t\in[0,+\infty),$
and its Mathematica animation is available [37].

Adding bound states
in our example results in global-in-time solutions containing
solitons. For example, by choosing
$$A=\bmatrix 1/2&-\sqrt{3}/2&0\\
\sqrt{3}/2&1/2&0\\
0&0&\kappa_1\endbmatrix,\quad B=\bmatrix 0\\
1\\
1\endbmatrix,\quad C=\bmatrix 2\gamma&2\epsilon&
c_1\endbmatrix,\tag 5.4$$
we get another explicit solution to (1.1) valid
for all $x\in[0,+\infty)$ and $t\in[0,+\infty).$
An explicit display of $u(x,t)$ corresponding to (5.4) is available in a Mathematica
file, but it takes many pages to display it; its animation with
$\epsilon=1/2,$ $\gamma=1/2,$ $\eta=1,$ $\kappa_1=2,$ and $c_1=3$ is also
available in the same Mathematica file [37]. The explicit global-in-time solution and its
Mathematica animation are also available [37] in a Mathematica file for the choices
$$A=\bmatrix 1/2&-\sqrt{3}/2&0&0\\
\sqrt{3}/2&1/2&0&0\\
0&0&\kappa_1&0\\
0&0&0&\kappa_2\endbmatrix,\quad B=\bmatrix 0\\
1\\
1\\
1\endbmatrix,\quad C=\bmatrix 2\gamma&2\epsilon&
c_1&c_2\endbmatrix,$$
which contains two solitons.

\vskip 10 pt

\noindent{\bf Acknowledgment}. The research leading to this article was
supported in part by the National Science Foundation under grant
DMS-0610494, the
Italian Ministry of Education and Research (MIUR) under COFIN grant no.
2004015437, and INdAM-GNCS.

\bigskip

\noindent {\bf REFERENCES}

\item{[1]}
J. L. Bona, S. M. Sun, and B. Y. Zhang, {\it
A non-homogeneous boundary-value problem for the
Korteweg-de Vries equation in a quarter plane,}
Trans. Amer. Math. Soc. {\bf 354}, 427--490 (2002).

\item{[2]} A. S. Fokas, {\it
Integrable nonlinear evolution equations on the half-line,}
Comm. Math. Phys. {\bf 230}, 1--39 (2002).

\item{[3]} N. J. Zabusky and M. D. Kruskal,
{\it Interaction of ``solitons" in a collisionless plasma and
the recurrence of initial states,} Phys. Rev. Lett. {\bf 15}, 240--243
(1965).

\item{[4]} M. J. Ablowitz and H. Segur, {\it
Solitons and the inverse scattering
transform,} SIAM, Philadelphia, 1981.

\item{[5]} M. J. Ablowitz and P. A. Clarkson, {\it Solitons, nonlinear
evolution equations and inverse scattering,} Cambridge Univ. Press, London,
1991.

\item{[6]} C. S. Gardner, J. M. Greene, M. D. Kruskal and R. M. Miura,
{\it Method for solving the Korteweg-de Vries equation,}
Phys. Rev. Lett. {\bf 19}, 1095--1097 (1967).

\item{[7]} V. A. Marchenko, {\it Nonlinear equations and operator algebras,}
D. Reidel Publishing Co., Dordrecht, 1988.

\item{[8]} V. B. Matveev and
M. A. Salle, {\it Darboux transformations and solitons,}
Springer-Verlag, Berlin, 1991.

\item{[9]} T. Aktosun, {\it Solitons and inverse scattering transform,}
In: D. P. Clemence and G. Tang (eds.),
{\it Mathematical studies in nonlinear wave propagation,}
Contemp. Math., Vol. {\bf 379},
Amer. Math. Soc., Providence, 2005, pp. 47--62.

\item{[10]} J. E. Colliander and C. E. Kenig, {\it
The generalized Korteweg-de Vries equation on the half line,}
Comm. Partial Differential Equations {\bf 27}, 2187--2266 (2002).

\item{[11]} T. Aktosun, M. Klaus, and C. van der Mee, {\it
Explicit Wiener-Hopf factorization for certain nonrational matrix functions,}
Integral Equations Operator Theory {\bf 15}, 879--900 (1992).

\item{[12]} H. D.Wahlquist and F. B. Estabrook, {\it
Prolongation structures and nonlinear evolution
equations,} J. Math. Phys. {\bf 16}, 1--7 (1975).

\item{[13]} B. G. Konopel'chenko, {\it
On exact solutions of nonlinear integrable equations via integral
linearising transforms and generalised B\"acklund-Darboux transformations,}
J. Phys. A {\bf 23}, 3761--3768 (1990).

\item{[14]} Nian Ning Huang, {\it
Darboux transformations for the Korteweg-de Vries equation,}
J. Phys. A {\bf 25}, 469--483 (1992).

\item{[15]} N. C. Freeman and J. J. C. Nimmo,
{\it Soliton solutions of the Korteweg-de Vries
and Kadomtsev-Petviashvili equations: the Wronskian technique,} Phys. Lett. A {\bf 95},
1--3 (1983).

\item{[16]} V. B. Matveev, {\it Generalized Wronskian formula for
solutions of the KdV equations: first applications,}
Phys. Lett. A {\bf 166}, 205--208 (1992).

\item{[17]} B. Carl and C. Schiebold, {\it
Ein direkter Ansatz zur Untersuchung von So\-li\-tonen\-glei\-chun\-gen,}
Jahresber. Deutsch. Math.-Verein. {\bf 102}, 102--148 (2000).

\item{[18]} Wen-Xiu Ma and Yuncheng You, {\it
Solving the Korteweg-de Vries equation by its bilinear form:
Wronskian solutions,}
Trans. Amer. Math. Soc. {\bf 357}, 1753--1778 (2005).

\item{[19]} R. Hirota, {\it The direct method in soliton theory,}
Cambridge University Press, Cambridge, 2004.

\item{[20]} A. Treibich and J. L. Verdier, {\it
Solitons elliptiques,}
Progr. Math., 88, The Grothendieck Festschrift,
Vol. III, Birkh\"auser-Boston, Boston, 1990,
pp. 437--480.

\item{[21]} F. Gesztesy and R. Weikard, {\it
Elliptic algebro-geometric solutions of the KdV and AKNS
hierarchies---an analytic approach,}  Bull. Amer. Math. Soc. (N.S.)
{\bf 35}, 271--317 (1998).

\item{[22]} F. Gesztesy, K. Unterkofler, and R. Weikard, {\it
An explicit characterization of Calogero-Moser systems,}
Trans. Amer. Math. Soc. {\bf 358}, 603--656 (2006).

\item{[23]} M. J. Ablowitz and J. Satsuma,
{\it Solitons and rational solutions of nonlinear evolution equations,}
J. Math. Phys. {\bf 19}, 2180--2186 (1978).

\item{[24]} V. A. Arkad'ev, A. K. Pogrebkov, and M. K. Polivanov,
{\it Singular solutions of the KdV equation and the inverse scattering method,}
J. Math. Sci. {\bf 31}, 3264--3279 (1985).

\item{[25]} V. B. Matveev, {\it
Positon-positon and
soliton-positon collisions: KdV case,} Phys. Lett. A {\bf 166}, 209--212 (1992).

\item{[26]} M. Kovalyov, {\it Basic motions of the Korteweg-de Vries
equation,} Nonlinear Anal. {\bf 31}, 599--619 (1998).

\item{[27]} I. Gohberg, M. A. Kaashoek, and A. L. Sakhnovich,
{\it Sturm-Liouville systems with rational Weyl functions: explicit
formulas and applications,} Integral Equations Operator Theory {\bf 30},
338--377 (1998).

\item{[28]} M. Jaworski, {\it
Breather-like solutions to the Korteweg-de Vries equation,} Phys. Lett. A {\bf 104},
245--247 (1984).

\item{[29]} Jian-qin Mei and Hong-qing Zhang, {\it
New soliton-like and periodic-like solutions for the KdV equation,}
Appl. Math. Comput. {\bf 169}, 589--599 (2005).

\item{[30]} T. Aktosun and M. Klaus, {\it
Chapter 2.2.4,
Inverse theory: problem on the line,}
In: E. R. Pike and P. C. Sabatier (eds.),
{\it Scattering,} Academic Press, London, 2001,
pp. 770--785.

\item{[31]} L. D. Faddeev, {\it Properties of the $S$-matrix of the
one-dimensional Schr\"odinger equation,} Amer. Math. Soc. Transl.
(ser.~2) {\bf 65}, 139--166 (1967).

\item{[32]} P. Deift and E. Trubowitz, {\it Inverse scattering on
the line,} Comm. Pure Appl. Math. {\bf 32}, 121--251 (1979).

\item{[33]} V. A. Marchenko, {\it Sturm-Liouville operators and
applications,} Birk\-h\"au\-ser, Basel, 1986.

\item{[34]} K. Chadan and P. C. Sabatier, {\it Inverse problems in
quantum scattering theory,} 2nd ed., Springer, New York, 1989.

\item{[35]} H. Bart, I. Gohberg, and M.A. Kaashoek, {\it Minimal
factorization of matrix and operator functions,} Birkh\"auser,
Basel, 1979.

\item{[36]} E. A. Coddington and N. Levinson, {\it Theory of ordinary
differential equations,} McGraw-Hill, New York, 1955.

\item{[37]} http://omega.uta.edu/$\sim$aktosun

\end